\documentclass[sigconf]{acmart}
\AtBeginDocument{%
  \providecommand\BibTeX{{%
    \normalfont B\kern-0.5em{\scshape i\kern-0.25em b}\kern-0.8em\TeX}}}

\usepackage{enumitem}
\usepackage{float}
\usepackage{soul}
\usepackage{xcolor}
\usepackage{etoolbox}
\usepackage{xcolor}

\newenvironment{compactquote}{%
  \par\smallskip
  \leftskip=0.5em
  \rightskip=0em
  \itshape\color{gray}%
  \parindent=0pt
  \noindent\ignorespaces
}{%
  \par\smallskip
}

\usepackage{subcaption}

\sethlcolor{white}
\begin{document}
\title{Analyzing LLM Usage in an Advanced Computing Class in India}

\author{Anupam Garg$^*$}
\email{anupam20555@iiitd.ac.in}
\affiliation{%
  \institution{IIIT Delhi}
  \city{New Delhi}
  \country{India}
  }

\author{Aryaman Raina$^*$}
\email{aryaman20034@iiitd.ac.in}
\affiliation{%
  \institution{IIIT Delhi}
  \city{New Delhi}
  \country{India}
}

\author{Aryan Gupta$^*$}
\email{aryan21314@iiitd.ac.in}
\affiliation{%
  \institution{IIIT Delhi}
  \city{New Delhi}
  \country{India}
  }

\author{Jaskaran Singh$^*$}
\email{jaskaran20306@iiitd.ac.in}
\affiliation{%
  \institution{IIIT Delhi}
  \city{New Delhi}
  \country{India}
  }

\author{Manav Saini$^*$}
\email{manav20518@iiitd.ac.in}
\affiliation{%
  \institution{IIIT Delhi}
  \city{New Delhi}
  \country{India}
  }

\author{Prachi IIITD$^*$}
\email{prachi20098@iiitd.ac.in}
\affiliation{%
  \institution{IIIT Delhi}
  \city{New Delhi}
  \country{India}
  }

\author{Ronit Mehta$^*$}
\email{ronit20539@iiitd.ac.in}
\affiliation{%
  \institution{IIIT Delhi}
  \city{New Delhi}
  \country{India}
  }

\author{Rupin Oberoi$^*$}
\email{rupin20571@iiitd.ac.in}
\affiliation{%
  \institution{IIIT Delhi}
  \city{New Delhi}
  \country{India}
}

\author{Sachin Sharma$^*$}
\email{chaitanya21559@iiitd.ac.in}
\affiliation{%
  \institution{IIIT Delhi}
  \city{New Delhi}
  \country{India}
}

\author{Samyak Jain$^*$}
\email{samyak21560@iiitd.ac.in}
\affiliation{%
  \institution{IIIT Delhi}
  \city{New Delhi}
  \country{India}
}

\author{Sarthak Tyagi$^*$}
\email{sarthak20540@iiitd.ac.in}
\affiliation{%
  \institution{IIIT Delhi}
  \city{New Delhi}
  \country{India}
  }

\author{Utkarsh Arora$^*$}
\email{utkarsh20143@iiitd.ac.in}
\affiliation{%
  \institution{IIIT Delhi}
  \city{New Delhi}
  \country{India}
}
 
\author{Dhruv Kumar}
\email{dhruv.kumar@iiitd.ac.in}
\affiliation{%
  \institution{IIIT Delhi and BITS Pilani}
  \city{New Delhi}
  \country{India}
}


\renewcommand{\shortauthors}{Anonymous et al.}

\begin{abstract}

This study examines the use of large language models (LLMs) by undergraduate and graduate students for programming assignments in advanced computing classes. Unlike existing research, which primarily focuses on introductory classes and lacks in-depth analysis of actual student-LLM interactions, our work fills this gap. We conducted a comprehensive analysis involving 411 students from a Distributed Systems class at an Indian university, where they completed three programming assignments and shared their experiences through Google Form surveys.

Our findings reveal that students leveraged LLMs for a variety of tasks, including code generation, debugging, conceptual inquiries, and test case creation. They employed a spectrum of prompting strategies, ranging from basic contextual prompts to advanced techniques like chain-of-thought prompting and iterative refinement. While students generally viewed LLMs as beneficial for enhancing productivity and learning, we noted a concerning trend of over-reliance, with many students submitting entire assignment descriptions to obtain complete solutions.
Given the increasing use of LLMs in the software industry, our study highlights the need to update undergraduate curricula to include training on effective prompting strategies and to raise awareness about the benefits and potential drawbacks of LLM usage in academic settings.

\end{abstract}





\maketitle
\def\thefootnote{$^*$}\footnotetext{These authors contributed equally to this work}\def\thefootnote{\arabic{footnote}}

\section{Introduction}

The recent emergence of advanced LLMs like GPT-3.5, GPT-4 \cite{openai_gpt-4_2023}, and LLama2 \cite{touvron2023llama} has revolutionized the field of artificial intelligence (AI). These models excel in a large variety of tasks including text generation, complex input understanding, and code generation.



This has sparked significant research into integrating LLMs in computing education \cite{becker2023ProsAndCons, Malinka2023Security, Daun2023Software, Denny2023CopilotCS1, Finnie-Ansley2022CS1, wermelinger2023Copilot, Savelka2023MCQAndCode, Reeves2023Parsons, finnie-ansley2023CodexCS2, Ouh2023Java, Cipriano2023GPT-3OOP, sarsa2022AutoGenerate, Leinonen2023CodeExplanation, Leinonen2023ExplainError, MacNeil2023CodeExplain, Balse2023Feedback}, with studies focusing on learning outcomes, personalization, and addressing concerns about academic integrity and critical thinking. These  studies have also discussed the challenges and opportunities educators and students face as they adapt to LLMs’ ability to generate source code from natural language descriptions.
Most of the existing research on LLMs in computing education has focused on introductory programming classes, covering applications such as code generation \cite{Denny2023CopilotCS1, Finnie-Ansley2022CS1, wermelinger2023Copilot, Savelka2023MCQAndCode, Reeves2023Parsons, finnie-ansley2023CodexCS2}, explanation of code \cite{MacNeil2023CodeExplain, Leinonen2023CodeExplanation}, debugging code \cite{Leinonen2023ExplainError} and development of supportive tools for students \cite{kazemitabaar2024codeaid, codehelpusinglarge}.


However, there is a notable gap in our understanding of the role of LLMs in advanced computing classes, particularly those designed for senior undergraduate and postgraduate students, where the focus shifts from writing code from scratch to designing and constructing complex systems and troubleshooting. Given the limited research on the applicability of LLMs in advanced computing classes, our study aims to address these gaps by analyzing interactions between students and LLMs in a Distributed Systems class at an Indian university. 

In the context of programming assignments within advanced computing classes, this report aims to achieve the following objectives:
\begin{itemize}
    \item \textbf{Objective 1:} Identify the usage patterns of LLM-based tools among students, focusing on the types of queries and styles of prompts used.
    \item \textbf{Objective 2:} Evaluate the effectiveness of responses from various LLMs in assisting students with their programming assignments.
    \item \textbf{Objective 3:} Understand students' perceptions on the influence of LLMs on their learning outcomes and productivity.
\end{itemize}
To investigate these objectives, we analyzed data from a Distributed Systems class where students were assigned three take-home programming assignments. These assignments covered key distributed systems topics, including communication libraries (e.g., gRPC, ZeroMQ, RabbitMQ), consensus protocols (e.g., Raft), and distributed computing frameworks (e.g., MapReduce).  Students were allowed to work in groups of up to three members, use LLMs as desired, and were required to submit both their code and a Google Form detailing their experiences with LLMs, along with the chat transcripts of their interactions.

Our data analysis approach involved summarizing the student responses to objective questions in the Google Form, presenting them in tabular format where possible, and conducting thematic analyses of both the open-ended question responses and the chat transcripts from the students' interactions with the LLMs.



The findings of our study reveal that students actively utilized LLMs for various purposes, including code generation, debugging, conceptual queries, and test-case generation, employing diverse prompting strategies such as contextual prompting, targeted querying, iterative refinement, problem decomposition, and advanced techniques like chain-of-thought prompting.
While students generally viewed LLMs positively for enhancing productivity and learning, we observed a concerning trend of over-reliance, with many students inputting entire assignment descriptions to seek complete solutions. Given the increasing adoption of LLMs in the software industry, our study highlights the need to update undergraduate curricula to include training on effective prompting strategies and to raise awareness about both the benefits and potential drawbacks of LLM usage in academic settings. Our research adds to the ongoing dialogue about the integration of AI tools in higher education and provides insights that could shape how we teach advanced computing classes in the future.

\section{Related Work}
There have been numerous studies that have explored the impact, challenges, and also opportunities that come with these new technologies \cite{becker2023ProsAndCons, Malinka2023Security, Daun2023Software, Denny2023CopilotCS1, Finnie-Ansley2022CS1, wermelinger2023Copilot, Savelka2023MCQAndCode, Reeves2023Parsons, finnie-ansley2023CodexCS2, Ouh2023Java, Cipriano2023GPT-3OOP, sarsa2022AutoGenerate, Leinonen2023CodeExplanation, Leinonen2023ExplainError, MacNeil2023CodeExplain, Balse2023Feedback}. 

\noindent\textbf{Integration of LLMs in Higher Education: }
Kasneci et al. \cite{chatgptforgood} and Becker et al. \cite{becker2023ProsAndCons} have done a deep dive into the benefits and challenges of using LLMs in education settings and emphasize the need for adaptation and ethical considerations. 

\noindent\textbf{Educator and Student Perspectives: } 
Lau et al. \cite{frombanittill} and Sheard et al. \cite{instructorperceptionsofai} presented a multi-institute interview-based study on instructors' perspectives regarding AI-based tools in education. 
Joshi et al. \cite{withgreatpower} and Budhiraja et al. \cite{itsnotlikejarvis} discuss the viewpoints of both students and professors on the influence of LLMs on education. 
Kazemitabaar et al. \cite{kazemitabaar2024codeaid} also analyze such conversations. However, they focus on tools that are specifically designed not to reveal complete code solutions in the context of introductory programming classes. 


Overall, these papers reveal a spectrum of opinions on the topic, ranging from cautious acceptance to enthusiasm. 


\noindent\textbf{Code Generation using LLMs: } 
Most of the existing studies have evaluated the effectiveness of LLMs for generating code in introductory programming exercises \cite{Finnie-Ansley2022CS1, wermelinger2023Copilot, thrilledbyyourprogress, Savelka2023MCQAndCode, Ouh2023Java, gpt3vsobject, Malinka2023Security, Daun2023Software, chatgptintheclassroom}. Our work differs from these studies as we focus on the usage of LLM-based tools in an advanced computing class (Distributed Systems).


\noindent\textbf{LLMs for supporting students: }
Existing studies have explored various aspects of how LLMs can support students. These include designing LLM-based tools \cite{tspace}, balancing student and educator needs \cite{codehelpusinglarge, kazemitabaar2024codeaid}, enhancing programming error messages with LLMs \cite{Leinonen2023ExplainError}, investigating high-precision feedback for programming syntax errors using LLMs \cite{tungphung2023international}, generating code explanations through LLMs \cite{MacNeil2023CodeExplain}, and comparing code explanations generated by students versus those produced by LLMs \cite{Leinonen2023CodeExplanation}.
\noindent\textbf{Prompting Strategies: }
Denny et al. \cite{Denny2023CopilotCS1, denny2023promptly} and Reeves et al. \cite{Reeves2023Parsons} analyzed the impact and usefulness of prompting techniques for solving programming problems in introductory programming classes. In comparison, our paper contributes by examining their awareness and implementation of these techniques, the extent of usage, and the specific types of techniques employed by students for solving programming problems in advanced computing class.

\noindent\textbf{Key Differences from Existing Studies} Our study differs from these existing studies in three aspects: (1) we focus on an advanced computing class (Distributed Systems), (2) we analyze the raw conversations of students with publicly available LLMs, (3) students are allowed to use LLMs in any way they wish, without any restrictions.



\section{Methodology}
\subsection{Details related to class and programming assignments}

\noindent \textbf{Demographics:} Our study focused on Undergraduate (UG) students in their junior and senior years and graduate students pursuing Master’s and Ph.D. in computer science and related disciplines at a tier-one Indian University enrolled in the class "Distributed Systems". The class comprised a total of 411 students. Of those students who were enrolled, there were 161 junior-year UG students, 232 senior-year UG students, 16 Master’s students, and 2 Ph.D. students. There were a total of 366 male students and 45 female students. There was only 1 section in this class, with 1 instructor and 15 Teaching Assistants (TAs). 

\noindent \textbf{Class Description and Components:} The "Distributed Systems" class aimed to provide students with a thorough understanding of core distributed systems concepts and develop their system design and implementation skills. Over 16 weeks, the class covered key topics like communication, coordination, consensus, and fault tolerance, along with real-world case studies such as MapReduce/Hadoop, Raft, Google Spanner, and Google Borg. The evaluation comprised quizzes (10\%), mid-semester exam (20\%), end-semester exam (30\%), and three take-home programming assignments (40\% total, split into 15\%, 15\%, and 10\%). This study concentrates on the programming assignments.

\noindent \textbf{Assignment Description:} Students completed three assignments over the course, working in groups of up to three (151 groups total). Assignment durations were 17, 24, and 13 days respectively, with relevant concepts covered in lectures beforehand. Students could use LLMs and their preferred programming language (Python, C++, Java). Assignments were distributed and submitted via the class management portal, with students submitting a zip file containing all deliverables. During the evaluation, students were expected to demonstrate the correctness and functionality of their program and respond to questions posed by the TA.\\
\noindent\textbf{Assignment 1} focused on developing three distinct distributed client server applications: online shopping platform using gRPC, group messaging platform using ZeroMQ, publisher-subscriber application like Youtube using RabbitMQ. \noindent\textbf{Assignment 2} focused on developing a distributed fault tolerant data storage application using a modified version of Raft, a popular consensus algorithm. \noindent\textbf{Assignment 3} focused on the development of a MapReduce-like application to execute K-means clustering in a distributed manner.

\noindent Complete assignment details are available on \href{https://github.com/researchtrack/Distributed-Systems-CSE-530}{GitHub} \cite{dscdassignment}.
\vspace{-1em}
\subsection{Study Design}


\noindent \textbf{Data Collection:}
At the time of code submission for each assignment, each student group was required to complete a form detailing their experiences with utilizing LLMs for solving the assignment.
Each group designated one member to submit a comprehensive survey on their collective LLM experience. The questionnaire delved into various aspects, including specific LLMs used, overall user experience, assignment completion times, and perceived impacts on productivity and learning outcomes. Students reported on the extent of LLM assistance, including the percentage of code generated, challenges encountered, and additional tools or websites utilized. The survey also collected suggestions for improvement and any additional comments. Importantly, students were required to upload transcripts of their LLM interactions, providing valuable insight into their problem-solving processes. A complete list of the questions present in the Google Form can be found on \href{https://github.com/researchtrack/Distributed-Systems-CSE-530}{GitHub} \cite{dscdassignment}.

It was not mandatory for the students to fill out the Google Form; however, completing this form for each assignment accounted for 10\% of that assignment's marks. This was done to encourage students to fill out the Google Form. During the evaluation process, the Teaching Assistant (TA) verified the correctness of the form that they were supposed to fill out. This weightage and TA verification minimized the likelihood of poor-quality data submissions. \\

\noindent \textbf{Data Analysis:}
The Google Form contained a mixed set of objective and subjective questions. For the objective questions, we employed straightforward statistical analysis. To extract meaningful insights from the open-ended questions and LLM conversation transcripts, we conducted a thorough thematic analysis. This qualitative approach helped us uncover patterns and trends in the students' responses and interactions. Researchers who carried out the analysis were subject matter experts in distributed systems.\\

\noindent {\textbf{1. Quantitative Analysis of Objective Data collected from Google Form:}}
We summarized the student responses to the objective questions in the Google Form and presented them in a tabular format.\\
\textbf{2. Thematic Analysis of Open-ended responses collected from Google Form:} 
For this analysis, the researchers split the open ended responses into two high level aspects, (1) the prompting strategies used for prompting, and (2) feedback, comments and suggestions for improving LLMs. For each aspect, we applied an inductive approach where the researchers involved in this task read the responses thoroughly, discussed and decided upon the themes and codes, and updated them iteratively, including feedback from the class instructor. The researchers decided upon 30 codes for Aspect 1 and 17 codes for Aspect 2.
Due to space constraints, we have made the full codebook, including themes, codes, and representative quotes, available on \href{https://github.com/researchtrack/Distributed-Systems-CSE-530}{GitHub} \cite{dscdassignment}.

\noindent\textbf{3. Thematic Analysis of the student conversations with the LLM:} 
The thematic analysis of the chat transcripts began with three raters reviewing conversations from three randomly sampled groups, representing 2\% of all assignments, to develop an initial codebook. This process involved a deductive approach, where predefined themes guided the initial coding. Following this, each rater coded the queries independently. During this process, they identified and discussed differences in their coding to achieve a consensus and enhance their understanding of each code. This collaborative effort led to the refinement of the codebook. Subsequently, 40 chats (consisting of 523 prompts) were randomly selected from a total pool of 1337 chats to compute inter-rater reliability using Fleiss' Kappa scores \cite{fleiss1971measuring}, which was determined to be 0.768. With a refined codebook (Table \ref{table:codebook}) and reliable inter-rater agreement, a detailed deductive thematic analysis was conducted on 213 chats (2317 prompts), with each rater analyzing an equal portion of the data. This thematic analysis aimed to uncover patterns, themes, and insights within the chat transcripts, providing a comprehensive understanding of the conversations.
\subsection{Ethical Considerations}
Ethical guidelines were followed throughout the data collection and analysis process to protect the privacy and confidentiality of the participants. Steps were taken to anonymize the data and adhere to relevant data protection regulations and institutional policies. Participants were adequately informed about the nature of the research and their participation. They were informed about the usage of the collected data and given an explanation of the project, with an outline of the objective and methods. Consent and necessary permissions were obtained from the students, and the data was anonymized.

\section{Evaluation}

\subsection{General Trends}
\noindent\textbf{Percentage of students using LLMs.} Out of 151 groups, responses were received from 146 groups for Assignment 1, 139 groups for Assignment 2, and 135 groups for Assignment 3. For Assignment 1, 145 groups (99.3\%) reported utilizing LLMs for solving the assignment. Similarly, 136 groups (97.8\%) reported using LLM tools for Assignment 2, and 134 groups (99.3\%) did so for Assignment 3.  

\noindent\textbf{LLM Tools Used by Students.} Students were probed about the particular LLM tools they used while solving the assignment. This was a multiple-select question as students often used more than one tool. Table \ref{table:llm_tools_used} shows that ChatGPT (GPT-3.5) was the most popular tool among students with 88.80\% students using it for solving the assignment. This is expected as ChatGPT was one of the earliest entrants in the world of publicly available LLM-tools and it is still available free-of-cost. Other popular tools were ChatGPT (GPT-4) (used by 26.20\% groups) and GitHub Copilot (used by 29.26\% groups). This also depicts that a good fraction of students (26.20\%) are ready to pay for using LLM-based tools.

\noindent\textbf{Time taken by students to solve the assignments.} Table \ref{table:time_taken} shows the mean and standard deviation of the total time taken by each student group to complete the three programming assignments. Any point which was not falling within 3 sigma (std dev) of mean was marked as an outlier.
\begin{table}[htb]
    \centering
    \begin{tabular}{|c|c|c|} \hline 
         \textbf{Assignment}&  \textbf{With Outliers}& \textbf{Without Outliers}\\ \hline 
         Assignment 1&  37.6$\pm$35.1& 29.3$\pm$16.0\\ \hline 
         Assignment 2&  45.5$\pm$41.6& 35.8$\pm$20.5\\ \hline 
         Assignment 3&  25.3$\pm$24.8& 21.7 $\pm$ 14.7\\ \hline
    \end{tabular}
    \caption{Average Time taken (in hours) by Students}
    \label{table:time_taken}
    \vspace{-10mm}
\end{table}
\subsection{Code generation by LLMs}
Table \ref{table:code_generation_percentage} shows the percentage of assignment code which the students reported generating using LLMs. 32.47 percent of students reported that LLMs generated 40\% to 60\% of their code. A significant portion (32.24\%) indicated having 20\%-40\% of their code generated by LLMs.  In contrast, fewer students fell into the categories of 0\%-20\% (20.47\%), 60\%-80\% (10.12\%) and 80\%-100\% (3.53\%) LLM-generated code. This indicates that LLMs are not able to generate the entire code and it requires considerable amount of student intervention to be able to use the LLM code for solving the programming assignment. This again seems to verify our observation that LLMs are increasing productivity (by providing some portion of code) while not hindering student learning (by making students write the remaining code) in advanced computing classes like Distributed Systems.

\subsection{Usage Patterns}
The Google Form had a multi-select question which asked the students to select the types of queries which they posed to the LLM. Table \ref{table:use_cases_llms} provides a summary of the student responses for this question. 90.47\% of the students reported using LLMs for generating code while 86.19\% reported using LLMs for understanding concepts. Some of the other common use-cases included "Debugging code" (84.03\%) and "Providing alternative solutions to tasks" (64.06\%). Less common uses included "Generating code documentation or comments" (56.91\%), "Brainstorming implementations" (59.77\%), "Generating code test cases" (34.75\%), "Providing insights and suggesting design patterns (46.9\%), and "Getting suggestions on optimization of code (54.06\%).  This suggests that students rely on LLMs for core coding tasks and basic problem-solving but less frequently use them for more advanced activities such as designing solutions for given problems.

Similar results were found in the thematic analysis of the chat transcripts. The main findings from the thematic analysis (in terms of types of queries posed by students) are presented below:

\noindent\textbf{Code generation:}
A significant portion of conversations (37.09\% ; 9.71\% of prompts) involved students copying assignment descriptions to get direct solutions. Students also frequently used LLMs (13.29\% of prompts) to generate code for specific portions of assignment, with some providing detailed instructions or pseudo-code. In 15.45\% of the total conversations, students sought specific syntax help, such as creating a for-loop in JavaScript. Additionally, 16.36\% of interactions involved students asking for minor code adjustments or combining independently developed code from team members.

\noindent\textbf{Code Review and Debugging:} 
A prominent portion (11.52\%) of the prompts involved students providing error stack traces. Furthermore, 7.59\% of prompts involved students submitting functional code for the LLM to review, seeking identification of potential bugs or poor coding practices. This also included scenarios where students described undesired code behavior (e.g., "Centroids not getting updated" or "Code not being terminated using Ctrl+C") and asked the LLM to identify and rectify problematic code sections.

\noindent\textbf{High-level assistance and conceptual queries:} 
Students seldom (2.67\% of prompts) used LLMs to discuss high-level implementation strategies, including brainstorming code structure and evaluating architectural suitability (e.g., Pub-Sub models). This low usage likely stemmed from detailed assignment descriptions providing clear implementation guidelines. More commonly (5.43\% of prompts), students utilized LLMs to better understand algorithms, asking questions like "explain k-means algorithm" or "how does Raft handle network partitions", thus supplementing in-class instruction.


\noindent\textbf{Code Explanation:}
LLMs were frequently used for code explanation, accounting for 5.56\% of prompts. Students primarily requested explanations of entire code blocks, though some inquired about specific line functionalities. They also asked LLMs to identify where certain features were implemented and to assess the code's ability to handle specific edge cases.

\noindent\textbf{Generate code documentation:} 
Generating code documentation accounted for 0.94\% of the prompts. This may be partly because the assignment did not specifically ask for giving well-documented code. Students mostly used LLMs for generating README files, a task required only once the assignment is completed. \\
\noindent\textbf{Setup and configuration assistance:}
LLMs were utilized for setup-related queries, accounting for 5.74\% of analyzed prompts. Students sought assistance with library installations and Google Cloud environment configuration. A frequent request involved generating terminal commands for gRPC code, which students could directly copy and paste due to the conversation's existing knowledge of the codebase's directory structure.
\vspace{-1em}
\subsection{Prompting Strategies}

Our analysis of student responses to the Google Form revealed various prompting strategies used when interacting with LLMs. A total of 366 instances of prompting strategies were recorded from the student responses. Contextual prompting emerged as the most prevalent strategy, with 19.67\% of students providing detailed assignment information, including pseudocode and theoretical background. Targeted querying was also common, with 11.48\% combining specific requests and direct questions. Iterative refinement was frequently employed, with 16.94\% including debugging, iterative prompting, and feedback loops, allowing students to progressively improve their results. A code-centric approach was popular among students, with 17.49\% sharing existing code or pseudocode with the LLM and requesting improvements, explanations, or extensions. Some students demonstrated knowledge of more advanced prompting techniques, with 7.65\% in total of Chain of Thought, Zero-Shot, Few-Shot, and Role-Based prompting. Task decomposition was observed in 5.74\%, where students broke down complex problems and used incremental construction approaches. Simplification strategies were less common but still present, with 2.73\% of students explaining concepts in layman's terms or using simplified scenarios. Notably, a significant number of students (10.38\%) reported using no specific strategy, instead opting for a more natural, conversational style of interaction with the LLMs. This thematic analysis, as mentioned in Section 3, provides insights into the diverse approaches students took when engaging with LLMs for their assignments.

\begin{compactquote}
``We tried breaking the problem statement into smaller problems since we have noticed that whenever we give a bigger and more complex problem it usually gives a response which is not useful, so we tried giving smaller instructions and smaller problems so that it can work on it.'' [Student Group A]
\end{compactquote}

\begin{compactquote}
``We tried both zero-shot and one-shot prompting techniques. There were some parts of the assignment which were aided with an example, in those places, the LLM was able to understand the task better. It also worked better on being provided the pseudocode. However, in case of zero-shot where we only gave it instructions in natural language and the code to build upon, it did not produce perfect results.'' [Student Group B]
\end{compactquote}

\subsection{Helpfulness of LLM Responses}
The Google Form collected student ratings of their LLM experience on a Likert Scale (ranging from "Excellent" to "Very Poor"). As shown in Table \ref{table:student_experiences}, 10.95\% rated it "Excellent," 23.82\% "Very Good," 35.94\% "Good," 22.13\% "Fair," and 6.23\% "Poor," with only 0.94\% reporting "Very Poor." Overall, a large majority (75\%) of students had a positive experience with LLMs.


Similar results were found in our thematic analysis of student conversations with LLMs. A vast majority (96.977\%) of LLM-generated responses were useful, providing relevant fixes and useful code, with students needing minor follow-ups for assignment specifications. However, some LLM responses were contextually incorrect, providing incomplete or incompatible code despite being instructed for complete solutions. Occasionally, students received repetitive errors after applying suggested changes. Rarely, LLMs disregarded context entirely, such as generating Flask code when queries were about RabbitMQ.

\subsection{Challenges and Recommendations}

Based on the analysis from the Google Form, Table \ref{table:challenges} reveals that 75.69\% of students identified "Difficulty in getting relevant or accurate responses from the LLM" as the most significant challenge. Additionally, 56.42\% of students reported "Difficulty distinguishing between correct and incorrect solutions provided by the LLM." This indicates that while LLMs may generate seemingly correct code, it can still be functionally flawed. Another  challenge was "Integrating the LLM-generated code with existing code," which 52.38\% of students reported. This suggests that LLMs are not yet capable of producing complete code for complex classes , highlighting that, in advanced computing classes, student learning may not be significantly impacted by the presence of LLMs.

\noindent Continuing from the previous analysis, feedback from 136 subjective student responses revealed several issues with LLMs. Problems included receiving incorrect answers (2.21\%), encountering hallucinations (0.74\%), excessive back-and-forth interactions (1.47\%), and limitations in handling complex tasks (11.03\%). Concerns about over reliance on LLM tools were also noted (4.41\%).
A significant portion of students (19.12\%) desired LLMs with better context understanding and improved instruction-following capabilities. Many (5.88\%) wanted tools that could maintain larger conversational contexts and provide more interactive feedback. They sought LLMs that not only generate accurate code but also offer specific guidance on improving their work (6.62\%). There was also a demand for better error debugging and diverse solutions (7.35\%) to enhance coding skills. Personalization was another key request (10.29\%), with students wanting LLMs tailored for educational purposes and efficient use of subject knowledge bases (7.35\%).


\begin{compactquote}
``They should generate a structure or flow, along with providing the basic commands like compile commands and should show the boilerplate code and tell about each component in it in detail. That way, we will get a clear image of the code, and maybe in the future, our dependence on these models for checking small errors may reduce.'' [Student Group C]
\end{compactquote}
\begin{compactquote}
``There is a layer of abstraction between the intent of the user and the answer generated by various LLMs. There is a need of a specific LLM for one particular task, here it can be code generation in Python. One can explain the concept of various libraries threading and Google Cloud Services.'' [Student Group D]
\end{compactquote}

\noindent LLMs may currently be insufficient for generating complete code in complex classes, and their presence may not significantly alter the learning experience. However, this observation is speculative and more comprehensive research is needed to reach a definitive conclusion about the true impact of LLMs on student learning in these specialized fields.

\subsection{Impact on Learning and Productivity}
The Google Form survey also explored the perceived impact of LLMs on both productivity and learning. 
A significant majority (72.71\%) believed LLMs enhanced their productivity, with only a small portion reporting no effect (7.53\%) while the rest (18.5\%) remained undecided. When being asked whether the students perceived a negative impact of LLMs on their learning, majority of students (59.06\%) believed the usage of LLMs in the assignment caused no impact to their learning, a good fraction (23.53\%) were unsure and a small fraction (13.88\%) perceived a negative impact on their learning.
While students perceive LLMs as tools that enhance productivity without adversely affecting their learning outcomes, further research is necessary to validate this perceived effect and ensure its accuracy.

\begin{table}[t]
	\small
	\vspace{-1em}
    
	\begin{tabular}{|p{5cm}|p{2cm} |} 
		\hline
		\textbf{LLM Used}& \textbf{\% Population}\\
		\hline
		ChatGPT(3.5)& 88.80\%\\ 
		\hline
        ChatGPT (GPT-4)& 26.20\%\\ 
		\hline
	  Google Bard / Gemini& 10.90\%\\
        \hline
        Microsoft Bing & 7.13\%\\
        \hline
        GitHub Copilot& 29.26\%\\\hline
	\end{tabular}
	\caption{\textbf{LLMs Used by Students}}
	\label{table:llm_tools_used}
	\vspace{-2em}
\end{table}

\begin{table}[t]
	\small
	\vspace{-1em}
    
	\begin{tabular}{|p{5cm}|p{2cm} |} 
		\hline
		\textbf{Students' experience with LLMs}& \textbf{\% Population}\\
		\hline
		Excellent& 10.95\%\\ 
		\hline
        Very Good& 23.83\%\\ 
		\hline
	  Good& 35.94\%\\
        \hline
        Fair& 22.14\%\\
        \hline
        Poor& 6.23\%\\\hline
 Very Poor&0.94\%\\\hline
	\end{tabular}
	\caption{\textbf{Students' experience with LLMs}}
	\label{table:student_experiences}
	\vspace{-7mm}
\end{table}
    
    
\begin{table}[t]
	\small
    
	\begin{tabular}{|p{5cm}|p{2cm} |} 
		\hline
		\textbf{Percentage of code generated by LLMs}& \textbf{\% Population}\\
		\hline
		0\%-20\%& 20.47\%\\ 
		\hline
        20\%-40\%& 32.24\%\\ 
		\hline
	  40\%-60\%& 32.47\%\\
        \hline
        60\%-80\%& 10.12\%\\
        \hline
        80\%-100\%& 3.53\%\\\hline
	\end{tabular}
	\caption{\textbf{Percentage of Code generated by LLMs}}
	\label{table:code_generation_percentage}
	\vspace{-7mm}
\end{table}
\begin{table}[t]
	\small
    	\begin{tabular}{|p{5cm}|p{2cm} |} 
		\hline
		\textbf{Use cases of LLMs by students}& \textbf{\% Population}\\ \hline 
 Code Snippet Generation&90.47\%\\ \hline 
 Explanation of Concepts&86.19\%\\ \hline 
 Providing alternative solutions to tasks&64.06\%\\ \hline 
 Debugging Code&84.03\%\\\hline
		Generating code documentation or comments& 56.91\%\\ 
		\hline
        Brainstorming Implementations& 59.77\%\\ 
		\hline
	  Generating code test cases& 34.75\%\\\hline
 Providing insights and suggesting design patterns&46.9\%\\ \hline 
 Getting suggestions on optimisation of code&54.06\%\\ \hline
	\end{tabular}
	\caption{\textbf{Use cases of LLMs}}
	\label{table:use_cases_llms}
	\vspace{-5mm}
\end{table}
\begin{table}[t]
	\small
	\vspace{-1em}
    	\begin{tabular}{|p{5.3cm}|p{1.7cm} |} 
		\hline
		\textbf{Challenges}& \textbf{\% Population}\\ \hline 
 Difficulty in getting relevant or accurate responses&75.69\%\\ \hline 
 Potential misunderstandings and misconseptions&24.27\%\\ \hline 
 Integrating generated code with existing code&52.38\%\\ \hline 
 Generating code snippets that fit into specific contexts&43.56\%\\\hline
		Uncertainty about the tools' limitations& 37.12\%\\\hline
        Difficulty distinguishing between correct and incorrect solutions& 56.43\%
\\ 
		\hline
	  Limited documentation surrounding the tool& 17.04\%\\\hline
	\end{tabular}
	\caption{\textbf{Challenges faced by students}}
	\label{table:challenges}
	\vspace{-1em}
\end{table}

\begin{table}[t]
	\small
	\vspace{-1em}
    
	\begin{tabular}{|p{4mm}|p{2.2cm}|p{5cm}|} 
		\hline
		\textbf{No.} & \textbf{Code} & \textbf{Explanation} \\
		\hline
		1 & Error message & Explaining the error or directly pasting the error message \\ 
		\hline
        2 & Code review & Getting unexpected/incorrect output, asking to find bug in code, check code, asking to predict output of code \\ 
		\hline
	    3 & System design/ brainstorming & Asking the LLM to help with a problem at a higher level and the implementation steps \\
        \hline
        4 & Conceptual queries & Asking theoretical or conceptual doubts \\
        \hline
        5 & Syntactical queries & Asking about some specific function/syntax \\
        \hline
        6 & Modify existing code & Asking to modify some functionality, or combine some given code snippets \\
        \hline
        7 & Fresh code generation & Asking to give code according to some specifications or add new functionalities \\
        \hline
        8 & Direct Code Solution & Directly asking the LLM to give the code for the problem description \\
        \hline
        9 & Explain Code & Getting an explanation for the selected code \\
        \hline
        10 & Comment / documentation & Asking to generate comments in code or create description/ readme file \\
        \hline
        11 & Setup based & Related to installation of libraries, help in networking issues, terminal commands, configuring environment, etc. \\
        \hline
        12 & Others & Follow-ups, response regeneration, etc. \\
        \hline
	\end{tabular}
	\caption{\textbf{Codebook for thematic analysis of student conversations with LLMs}}
	\label{table:codebook}
	\vspace{-11mm}
\end{table}

\section{Discussion}

Our study revealed an overwhelming adoption of LLMs among students, with 98\% using them for assignments \cite{chatgptforgood, becker2023ProsAndCons}. Many found LLMs helpful in overcoming coding challenges and grasping complex concepts, though the extent of reliance varied widely \cite{withgreatpower, itsnotlikejarvis, frombanittill}. Students employed diverse prompting strategies in their LLM interactions, highlighting the importance of education on effective LLM usage \cite{kazemitabaar2024codeaid}. Proper training could significantly enhance LLMs' effectiveness as educational tools \cite{chatgptintheclassroom}.

However, the utility of LLMs raises ethical concerns. Students often copied and pasted LLM-generated code, prompting questions about over-reliance and academic integrity \cite{chatgptforgood, becker2023ProsAndCons}. This behavior suggests a need to reassess learning outcomes and evaluation methods to ensure genuine learning. Future research should incorporate both short-term and long-term assessments to gauge LLMs' true impact on learning and knowledge retention.

\noindent The increasing use of LLMs in the software industry underscores the value of exposing students to these tools in a controlled educational environment. This exposure helps students understand both the benefits (like faster coding and idea generation) and limitations (such as potential errors) of LLMs, fostering critical thinking and problem-solving skills crucial for professional success.

Integrating LLMs into education necessitates a reevaluation of course curricula and learning outcomes \cite{Leinonen2023CodeExplanation, tspace, codehelpusinglarge}. Educators must design assignments and assessments that cultivate skills beyond those easily addressed by LLMs, such as deeper understanding, creativity, and critical thinking. Updating learning outcomes to reflect the evolving landscape of educational tools will better prepare students for academic and professional challenges.

In essence, while LLMs offer significant benefits in computing education, their integration requires careful consideration of ethical implications, effective usage strategies, and curriculum adaptations. This balanced approach will ensure that LLMs enhance rather than hinder genuine learning and skill development, preparing students for the realities of the modern tech industry while maintaining academic integrity.
\section{Conclusion}

Our study highlights both the potential and challenges of integrating Large Language Models into higher education. To maximize their benefits, we must address ethical concerns, mitigate over-reliance, and teach effective usage. This necessitates updating our educational frameworks to align with the evolving technological landscape and enhance learning experiences to prepare students for their careers. However, our understanding of LLMs' long-term educational impact remains incomplete. Further research is needed to develop best practices for academic use. As we navigate this new frontier, ongoing collaboration between educators, researchers, and industry professionals will be key to shaping a balanced integration of AI tools in higher education.

\bibliographystyle{ACM-Reference-Format}
\bibliography{chatgpt-1}

\appendix

\end{document}